\documentclass{xray_symp}
\usepackage{graphics}
\def\G{$\Gamma_{\rm x}$ } 

\def\ros{{\sl ROSAT }} 
\def\asca{{\sl ASCA }}

\def\approxlt{\mathrel{\hbox{\rlap{\lower.55ex \hbox {$\sim$}}
        \kern-.3em \raise.4ex \hbox{$<$}}}}
\def\approxgt{\mathrel{\hbox{\rlap{\lower.55ex \hbox {$\sim$}}
        \kern-.3em \raise.4ex \hbox{$>$}}}}
\begin{document}
\title {Properties of warm absorbers in active galaxies}
\author {Stefanie Komossa \and Henner Fink$\dagger$}
\institute{Max--Planck--Institut f\"ur extraterrestrische Physik,
 Giessenbachstra{\ss}e, 85740 Garching, Germany}
\authorrunning{St. Komossa, H. Fink}
\titlerunning{~~Properties of warm absorbers in active galaxies}
\maketitle 

\begin{abstract}

We present a study of the nature of warm absorbers
on the basis of \ros X-ray observations and photoionization calculations
carried out with the code {\em Cloudy}.
We focus on `non-standard' warm absorbers:
(i) the proposed {\em dusty} warm absorbers in IRAS\,13349+2438 and 
IRAS\,17020+4544, and possibly  
in  4C\,+74.26,  
(ii) we explore several scenarios to account for the recently observed
peculiar absorption features around $\sim$1.1 keV on the basis of detailed
photoionization models, and apply these to PG\,1404+226.

\end{abstract}

\section {Introduction} 

Absorption edges in the X-ray spectra of Seyfert galaxies
have been interpreted as the signature of ionized gas along the line of sight to the active nucleus. 
These so-called `warm absorbers' provide an important 
new diagnostic of the AGN central region. 
So far, they revealed their existence mainly in the soft X-ray spectral region. 
The physical state and location of the ionized material 
and its relation to other components of the active nucleus is still rather unclear. 
E.g., an outflowing accretion disk wind and 
various BLR related models have been suggested.

We performed a study of the properties of warm absorbers 
based on \ros (Tr\"umper 1983) X-ray observations and photoionization calculations
carried out with the code {\em Cloudy} (Ferland 1993).
Results on the warm absorbers in NGC\,4051, NGC\,3227, NGC\,3786,
Mrk\,1298 were presented earlier in Komossa \& Fink (1997a-d, respectively). 
Here, we focus on `non-standard' warm absorbers:
(i) we comment on the influence of dust on the X-ray absorption spectrum, apply 
the model of a dusty warm absorber to 
IRAS\,13349 and IRAS\,17020, and present 4C\,+74 as new candidate, 
(ii) we investigate several scenarios to account for the recently observed 
peculiar X-ray features around $\sim$1.1 keV on the basis of detailed
photoionization models, and apply these to PG\,1404+226.

\section {Warm absorbers with internal dust} 

\subsection{Influence of dust on the X-ray absorption structure}  

Recently, evidence has accumulated that some warm absorbers
contain significant amounts of dust. This possibility was 
first suggested by Brandt et al. (1996)
to explain the lack of excess X-ray {\em cold} absorption despite strong optical
reddening (hereafter referred to as `$N_{\rm opt}-N_{\rm x}$ discrepancy')
of the quasar IRAS 13349+2438.

\begin{figure}[bh]
\resizebox{\hsize}{!}{\includegraphics{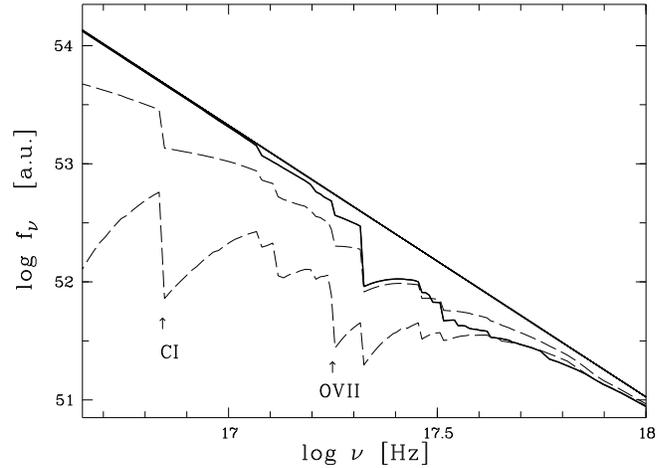}}
  \caption[]{Changes of the X-ray absorption spectrum in the presence of dust.
The thin straight line marks the intrinsic continuum, the fat line shows
a dust-free warm absorber.
The dashed lines correspond to the same model after inclusion of dust
and depleted gas-phase metal abundances.
The dust was depleted relative to the standard Galactic-ISM
mixture by factors of 10 (upper dashed curve) and 3 (lower
dashed curve).
A characteristic feature of (the graphite species of)
dust is the strong edge of neutral Carbon, labeled C\,I.
}
  \label{dust_seq}
\end{figure}

\begin{table*}  
 \footnotesize
  \caption{Properties of the warm absorbers from X-ray spectral fits
   and results from a single powerlaw fit for comparison.
}
\begin{center}
  \begin{tabular}{lllllll}
  \noalign{\smallskip}
  \hline
  \noalign{\smallskip}
     & \multicolumn{4}{l}{dusty warm absorber} & \multicolumn{2}{l}{single powerlaw} \\
         & \G~~~ & log $U$~~ & log $N_{\rm w}$~~ & $\chi^2_{\rm red}$~~~~~ &
                                                               \G~~~ & $\chi^2_{\rm red}$ \\
  \noalign{\smallskip}
  \hline
  \noalign{\smallskip}
    4C\,+74.26  & --2.2 & --0.1 & 21.6 & 1.0 & --1.4 & 1.0 \\
  \noalign{\smallskip}
    IRAS\,17020+4544  & --2.8 & ~\,0.7 & 21.6$^{(1)}$ & 0.8 &  --2.4 & 0.8 \\
     sil$^{(2)}$ & --2.4 & ~\,1.0 & 21.6$^{(1)}$ & 0.9  &       & \\
  \noalign{\smallskip}
    IRAS\,13349+2438  & --2.9 & --0.4 & 21.2$^{(1)}$ & 1.2 & --2.8 & 1.3 \\
      df$^{(3)}$  & --2.2 &   ~\,0.7 & 22.7 & 0.8 & & \\
  \noalign{\smallskip}
  \hline
  \noalign{\smallskip}
     \end{tabular}
  \label{tab1}

  \noindent{ \scriptsize
$^{(1)}$ fixed to the value $N_{\rm opt}$ determined from optical reddening ~
$^{(2)}$ silicate species of dust only ~ $^{(3)}$~dust-free
warm absorber for comparison }
\end{center}
\end{table*}

As we emphasized earlier (Komossa \& Fink, e.g. 1997a,b; Komossa \& Bade 1998)
and demonstrate here in Fig. \ref{dust_seq} the
influence of the presence of dust on the
X-ray absorption spectrum can be strong, and becomes drastic for high column densities $N_{\rm w}$.
Signatures of the presence of (Galactic-ISM-like) dust
are, e.g., a strong carbon edge in the X-ray spectrum,
and a stronger temperature gradient across the absorber
with more gas in a `colder' state.
Another interesting property is the increased sensitivity of dusty gas
to radiation pressure, which may drive strong outflows of the warm material.

\subsection{4C\,+74.26} 
4C\,+74 is a radio-loud quasar. In an analysis of \ros and \asca data,
Brinkmann et al. (1998) find an unusually flat soft X-ray \ros spectrum
(\G $\simeq -1.3~{\rm to} -1.6$; as compared to \G $\simeq -2.2$ typically seen in     
radio-loud quasars), a steeper \asca powerlaw (PL) spectrum, and evidence for the
presence of a warm absorber.  
Applying the model of a {\em dusty} warm absorber to the \ros spectrum
we get a successful spectral fit, with a steeper intrinsic
PL spectrum (now consistent with the \asca value and the general expectation for
radio-loud quasars),  and a column density $N_{\rm w}$ 
consistent with optical reddening (Tab. 1).
Alternatively, excess {\em cold} absorption of large column density
fits the \ros spectrum,
and better spectral resolution soft X-ray data are
needed to exclude a cold absorber. 

\subsection{IRAS\,17020+4544}   

\noindent A dusty warm absorber was suggested to be present in the 
NLSy1 galaxy IRAS\,17020 (Leighly et al. 1997b) on the basis of  
an $N_{\rm opt}$ - $N_{\rm x}$
discrepancy and the presence of an absorption
edge in the \asca spectrum. 

When fit by a single powerlaw, the \ros X-ray spectrum of IRAS 17020 
is rather steep (\G = $-2.4$). 
Checking whether (and under which conditions) a dusty warm absorber
fits the \ros X-ray spectrum, we find that an even steeper intrinsic 
spectrum is required to compensate for the `flattening effect' (cf. Fig. \ref{dust_seq})
of dust (Tab. 1); for details on IRAS 17020 see Komossa \& Bade (1998). 

The presence of a {\em dusty} warm absorber in IRAS\,17020
particularly adds to the spectral complexity in NLSy1 galaxies.
Whereas early NLSy1 models tried to explain their very {\em steep} 
observed X-ray spectra
by only one component (either a strong soft excess, or a warm absorber, or an intrinsically
steep spectrum), there is now evidence that 
often all three components   
are simultaneously present, and the additional presence of {\em dusty}
material partly compensates the `steepening effect' of the other three.

\subsection{IRAS\,13349+2438}

This quasar received a lot of attention, recently.
A detailed optical study was presented by Wills et al. (1992).  
In X-rays, the presence of a dusty warm absorber was suggested (Brandt et al. 1996). 
Brinkmann et al. (1996) detected changes in the \asca spectrum as compared to
the earlier \ros data;
the warm-absorption features remained present (Brandt et al. 1997). 
Here, we apply the model of a {\em dusty} warm absorber to the \ros X-ray
spectrum. Although repeatedly suggested, such a model
has not been fit previously (for first results see Komossa 1998).  
Given the potentially strong modifications of the X-ray absorption spectrum
in the presence of dust,  
it is important to  scrutinize whether a dusty warm absorber is consistent with
the observed X-ray spectrum. 
Since some strong
features of dusty warm absorbers appear outside the \asca sensitivity range, 
\ros data are best suited for this purpose; we used the two pointed PSPC observations
of Jan. 1992 and Dec. 1992 (P-2 hereafter).  

In a first step, we fit a dust-{\em free} warm absorber (as in Brandt et al. 1996,
but using the additional
information on the hard X-ray powerlaw available from the ASCA observation,
$\Gamma_{\rm x}^{\rm 2-10 keV} \simeq -2.2$).
This gives an excellent fit with log $N_{\rm w}$=22.7 ($\chi^2_{\rm red}$ = 0.84 for P-2).  
If this same model is re-calculated by fixing $N_{\rm w}$ and
the other best-fit parameters but {\em adding dust} to the warm absorber
the X-ray spectral shape is drastically altered and the data can not be
fit at all ($\chi^2_{\rm red}$ = 150). This still holds if we allow for
non-standard dust, i.e., selectively exclude either the graphite or silicate
species.

It has to be kept in mind, though, that the expected column derived
from optical extinction is less than the X-ray value of $N_{\rm w}$ determined
under the above assumptions. Therefore, in a next step, we
allowed all parameters (except \G) to be free and checked, whether a dusty
warm absorber could be successfully fit at all.
This is not the case (e.g., if $N_{\rm w}$ is fixed to log $N_{\rm opt}$ = 21.2
we get $\chi^2_{\rm red}$ = 40).

The bad fit results can be partially traced back to the `flattening' effect of dust.
In fact, if we allow for a steeper intrinsic powerlaw spectrum, with \G $\simeq -2.9$ 
much steeper than the \asca value, 
a dusty warm absorber with $N_{\rm w}=N_{\rm opt}$ fits the \ros 
spectrum well ($\chi^2_{\rm red}$ = 1.2, Tab. 1).
We also analyzed the \ros survey data and find the same   
trends.   
At present,
there are several possible explanations for the {\sl ROSAT}-\asca spectral differences:
(i) variability in a {\em two}-component warm absorber, (ii) variability in
the intrinsic powerlaw, or (iii) remaining {\sl ROSAT}-\asca
inter-calibration uncertainties.

\section{Peculiar 1.1 KeV absorption}    

Recently, several cases of spectral complexity around 1.1 keV
have been reported (e.g., Hayashida 1997). Among these is PG\,1404+226.  
Its  \ros high-state and \asca spectrum 
show evidence for unexpectedly strong 1.1 keV absorption
(Ulrich \& Molendi 1996, Comastri et al. 1997).
If interpreted in terms of blueshifted oxygen,
exceptionally high outflow velocities are implied (e.g., Leighly et al.\,1997a);
an exciting possibility.

Here, on the basis of detailed photoionization modelling of the absorbing
material under various conditions, we explore several 
scenarios to account for the 1.1 keV absorption
{\em without} invoking relativistic outflow of the warm absorber (WA):

\begin{figure} 
 \resizebox{\hsize}{!}{\includegraphics{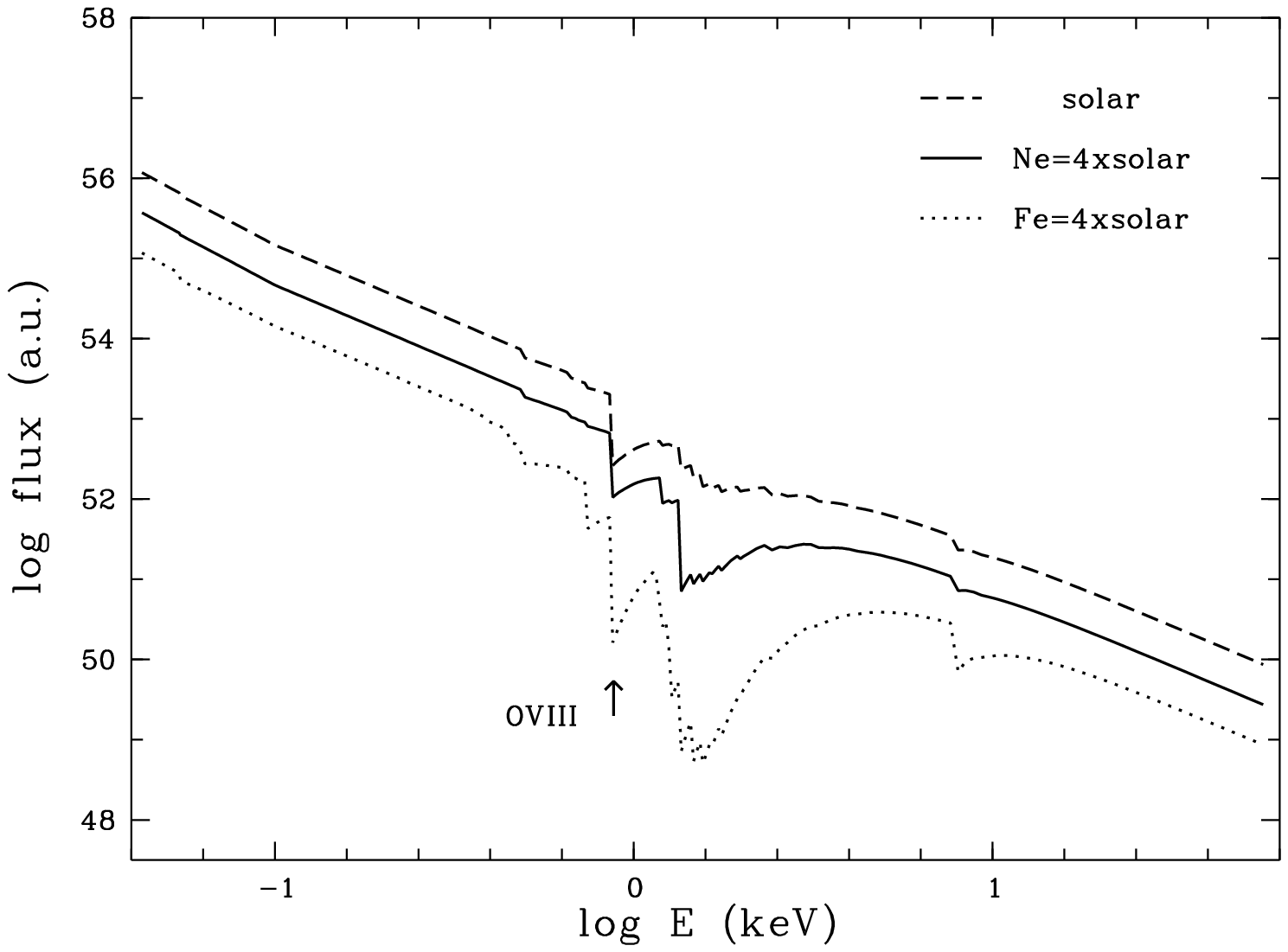}}

\vspace*{0.25cm}

 \resizebox{\hsize}{!}{\includegraphics{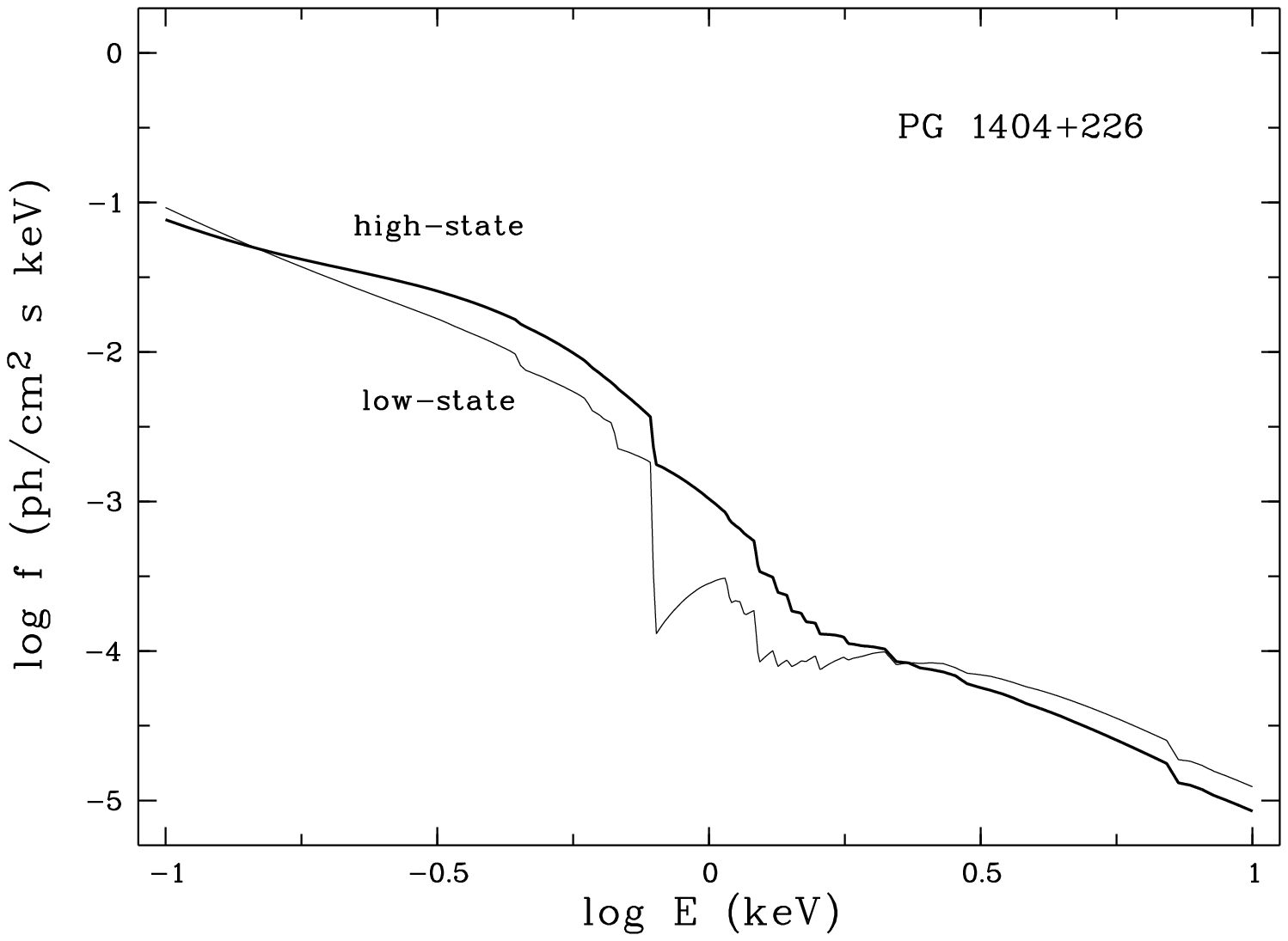}} 
 \caption[mod]{Upper panel: Changes in absorption structure in case of non-solar abundances.
The upper curve (dashed) gives a model with solar abundances; oxygen absorption
dominates over Ne-K and Fe-L. The next curve (solid line) corresponds to a model
with an abundance of Ne = 4$\times$solar; a strong neon edge is seen to emerge,
now dominating over oxygen. The lower curve (dotted) represents a model with 
overabundant iron of Fe=4$\times$solar; strong Fe-L edges are present in this case.
{\em All} other parameters than the abundances were fixed -- for presentation purposes,
the curves were shifted along the ordinate by 0.5 in log $f$.
Lower panel: Modified soft X-ray continuum, consisting of a powerlaw plus black body (bb)
both {\em incident} on the warm absorber, and changes of the X-ray spectrum 
after increasing the bb contribution.
The models plotted correspond to the best-fits within this prescription 
to the \ros data of PG\,1404.   
}
 \label{mod}
\end{figure}

\subsection {WA with contribution from emission and reflection}
Models of high ionization parameter $U$ and/or with the emission and reflection
component added to the observed spectrum, were calculated for a covering factor of 0.5. 
In application to PG 1404, fitting such a model 
improves
the quality of the fit, but the high-state data are still not
well matched.

\subsection {WA with non-solar O/Ne ratio} 
One way to clearly change the depth of individual absorption edges, and
particularly to make the neon absorption dominate over oxygen in strength,
is a deviation from solar abundances (Fig. \ref{mod}), of either
overabundant neon or underabundant oxygen. 

Several deviation factors were studied between an abundance of
up to O = 0.2 $\times$ solar and up to Ne = 4 $\times$ solar.
These models strongly improve the quality of the fit to PG\,1404
up to acceptable values (Tab. 2).
A potential problem for this model description, besides a difficulty 
to explain strong deviations of O/Ne from the solar value in terms
of nucleosynthesis, is the width of the X-ray absorption feature and
the location of the deepest edge. 
Always, several ionization stages of neon coexist, leading to a {\em broad}
absorption structure.     
Further, in order to weaken sufficiently the oxygen absorption, a rather
high ionization parameter is required with the consequence that 
the deepest neon edges are those of highly ionized species, around
1.36 keV, instead of 1.1 keV (Fig. \ref{mod}).  

   \begin{table*} 
     \caption{\small X-ray spectral fits to the high-state (HS) and low-state (LS)
             \ros data of PG 1404.
              \G was fixed to $-1.9$.
              Errors in $U$, $N_{\rm w}$
              are about a factor 2--3.
                                                }
     \label{fitres}
 \begin{center}
      \begin{tabular}{ccccl}
      \hline
      \noalign{\smallskip}
        state & log $U$ & log $N_{\rm w}$ &
                            $\chi^2_{\rm red}$ & model \\
      \noalign{\smallskip}
      \hline
      \noalign{\smallskip}
      HS/LS & & & 4.1/2.0 & single PL \\
     HS/LS & 1.0/0.7 & 23.7/23.4 & 2.5/0.9 & standard warm absorber \\
     HS/LS & 1.0/0.7 & 23.8/23.5 & 2.2/0.9 & emission+reflection added \\
     HS/LS & 0.7/0.3 & 23.3/22.9 & 1.3/0.9 & Ne = 4$\times$solar abundance \\
     HS/LS & 0.6/0.4 & ~~~23.1/23.1$^{(1)}$ & 1.2/1.0 & additional 0.1 keV soft excess$^{(2)}$\\
      \noalign{\smallskip}
      \hline
  \end{tabular}

\noindent{\scriptsize $^{(1)}$ fixed to value derived for HS; $^{(2)}$ abundances reset to solar
}
\end{center}
   \end{table*}

\subsection {Additional soft-excess {\em incident} on WA} 
Motivated by the \asca evidence for soft excesses in some NLSy1s,
a sequence of models was calculated with an additional hot
black body component of T = 0.1 keV. This component was included
in the ionizing SED that illuminates the absorber,
i.e. the change in
ionization structure of the warm material was self-consistently
calculated.  
This causes a complex spectral shape in the 1 keV region, with 
the down-turning soft-excess and some Ne-K and Fe-L absorption 
contributing to the X-ray features which then sensitively depend
on the strength of the soft excess (Fig. \ref{mod}).      
(Note that, since the bb is {\em incident} on the warm material, and not 
added afterwards as separate component, this causes a different ionization
structure and therefore soft X-ray spectral shape.) 
A successful description of both, high- and low-state \ros data
is possible, but again a rather broad absorption feature is predicted (Fig. 2).

In the latter two cases (3.2 and 3.3) 
the difference between high-state and low-state spectrum
could be explained by higher ionization parameter in high-state
or variable strength of the soft excess. 
The results are summarized in Tab. 2.  

Data of high spectral resolution will be needed to finally discriminate between
these models, the interesting alternatives of relativistic outflow  
(Leighly et al. 1997a) or 
high iron overabundance (Comastri et al. 1997), and further scenarios 
to account for the 1.1 keV features.

\section{Summarizing conclusions}

There is good evidence that several warm absorbers contain
dust as signified by multi-$\lambda$ evidence and successful X-ray spectral
fits. The influence of dust on the X-ray absorption structure can be strong.
{\em Dusty} warm absorbers are found to be consistent with the X-ray spectra of 
4C\,74, IRAS\,17020, and IRAS\,13349. Whereas in the first case, the flattening
effect of dust leads to consistency of the intrinsic X-ray spectrum with
the \asca shape and general expectations, in the latter case a rather steep
\ros spectrum is required. 

These several good cases of  {\em dusty} warm absorbers
suggest this component to be
common in all types of AGN.
Clear signatures of the presence of dust are a carbon edge at 0.28 keV
and an oxygen edge at 0.56 keV, not yet individually resolved by current X-ray
instruments. The study of these features with future missions will provide an
interesting approach to investigate dust properties in other galaxies.
But not all warm absorbers contain dust (cf. Komossa \& Fink 1997a, Komossa 1998).

Several scenarios to explain the recently observed
unusual 1.1 keV absorption feature were studied. 
Among these, non-solar O/Ne-ratio and the presence of an additional 
soft excess illuminating the warm gas 
are found to fit the \ros X-ray spectrum of PG\,1404.
Potential problems of these and alternative model descriptions are pointed out.  

In summary, warm absorbers display many facets and may account for a variety
of observational phenomena. Their detailed study
will certainly
be an important goal of future X-ray satellites like {\sl AXAF} and {\sl XMM}.

\begin{acknowledgements}
I gratefully remember {\bf Henner Fink} for 
introducing me to the work with X-ray data, for many discussions  
and helpful advice. 
Henner Fink passed away in December 1996. \\
It is a pleasure to thank Marie-Helene Ulrich and Andrea Comastri for stimulating discussions on
PG 1404, Gary Ferland for providing {\em Cloudy},
and Emmi Meyer-Hofmeister for her continuing kind interest. 
\end{acknowledgements}


\begin{thebibliography}{99}
\bibitem{} Brandt W.N. et al., 1996, MNRAS 278, 326; 1997, MNRAS 292, 407   
\bibitem{} Brinkmann W. et al., 1996, A\&A 316, L9; 1998, A\&A 330, 67  
\bibitem{} Comastri A. et al., 1997, in {\em X-Ray Imaging and Spectr. of
     Cosm. Hot Plasmas}, 
     F. Makino, K. Mitsuda (eds), 279  
\bibitem{} Ferland G.J., 1993, University of Kentucky, Physics Department, Internal Report 
\bibitem{} Hayashida K., 1997, in {\em Emission Lines in AGN},
           B.M. Peterson et al. (eds), ASP conf. ser. 113, 40 
\bibitem{}  Komossa S., 1998, in {\em Structure and Kinematics of Quasar Broad Line Regions},
M. Gaskell et al. (eds), ASP conf ser., in press 
\bibitem{}  Komossa S., Fink H., 1997a, A\&A 322, 719;
     ~1997b, A\&A 327, 483; ~1997c, A\&A 327, 555 
\bibitem{}  Komossa S., Fink H., 1997d, in {\em Accretion Disks -- New Aspects},
             E. Meyer-Hofmeister, H. Spruit (eds),  Lecture Notes in
             Physics 487, 250 
\bibitem{} Komossa S., Bade N., 1998, A\&A 331, L49
\bibitem{} Leighly K., et al., 1997a, ApJ 489, L25; ~ 1997b, ApJ 489, L137 
\bibitem{} Tr\"umper J., 1983, Adv. Space Res. 2, 241
\bibitem{} Ulrich-Demoulin M.-H., Molendi S., 1996, ApJ 457, 77 
\bibitem{} Wills B.J. et al., 1992, ApJ 400, 96
\end{thebibliography}
\end{document}